\documentclass[epsfig,12pt] {article}
\usepackage{graphics}
\usepackage{epsfig}

\def\n{\noindent}
\usepackage{epsfig}
\include{definition}
\begin{document}

\baselineskip .7cm

\author{Haribabu Arthanari $^{\ast}$, G. Wagner \thanks{Harvard Medical School},  Navin Khaneja \thanks{To whom correspondence may be addressed. Email:navin@hrl.harvard.edu. School of Engineering and Applied Sciences, Harvard University,
Cambridge, MA 02138.}}

\title{Heteronuclear Decoupling by Multiple Rotating Frame Technique}

\maketitle

\begin{center} {\bf Abstract} \end{center}

The paper describes the multiple rotating frame technique for designing modulated rf-fields,
that perform broadband heteronuclear decoupling in solution NMR spectroscopy. The decoupling is understood by performing a sequence
of coordinate transformations, each of which demodulates a component of
the Rf-field to a static component, that progressively averages the chemical shift and dipolar interaction. We show
that by increasing the number of modulations in the decoupling field, the ratio of dispersion in the chemical shift to
the strength of the rf-field is successively reduced in progressive frames.  The
known decoupling methods like continuous wave decoupling, TPPM etc, are special cases of this method and their
performance improves by adding additional modulations in the decoupling field. The technique is also expected to find use in
designing decoupling pulse sequences in Solid State NMR spectroscopy and design of various excitation, inversion and
mixing sequences.

\n
\vskip 3em

\section{Introduction}

Heteronuclear decoupling methods have a long history in NMR spectroscopy \cite{TPPM, Spinal-64, gan, eden, Yu, Takegoshi, XiX, Levitt_Rev, Shaka_Rev, Anderson_62, Anderson_63, Ernst_dec, Freeman_79, Levitt_82, Waugh_82, Waltz-16, Shaka_83, Shaka_85, Fujiwara_88, Zuiderweg_91, Eggenberger_92, McCoy_93, Fujiwara_93, Starcuk _94, Bendall _95, Skinner_96, Kupce_96, Bodenhausen_96, Geen_96, Geen_97} and {\em in vivo} applications \cite{Bottomley_89, Hurd_97, Shaka_in_vivo,  Graaf_in_vivo, Li_Yang_Shen, Chen_09}.
The goal of broadband heteronuclear decoupling sequences is to observe a spin $S$ by irradiating
spin $I$ that is coupled to $S$ in order to simplify the spectra and to increase the
signal-to-noise ratio. At the same time, the decoupling sequence should introduce only a minimal
amount of artifacts, such as decoupling sidebands. Furthermore, in order to avoid undesirable
sample heating or damage to the probe, the radio frequency (rf) power of the decoupling sequence
should be as small as possible. This is of particular importance in medical imaging
or {\em in vivo} spectroscopy of humans \cite{Bottomley_89, Hurd_97, Shaka_in_vivo,  Graaf_in_vivo, Li_Yang_Shen, Chen_09}.

The earliest heteronuclear decoupling methods were based on cw irradiation \cite{Anderson_62} and noise decoupling \cite{Ernst_dec}. Significantly improved decoupling sequences were found based on composite \cite{Levitt_Rev, Shaka_Rev, Freeman_79, Levitt_82, Waltz-16, Shaka_83, Shaka_85, Fujiwara_88} or shaped \cite{Starcuk _94, Bendall _95, Skinner_96, Kupce_96, Bodenhausen_96, Geen_97} inversion pulses in combination with
highly compensated cycles and supercycles \cite{Levitt_Rev, Shaka_Rev, Waltz-16, Levitt_8_super, Levitt_83, Tycko_85a, Tycko_85b,Tycko_90, Gullion_90, Lizak_91}.
Theoretical approaches  that have been used for the analysis and design of decoupling sequences include
average Hamiltonian \cite{Waugh_82, AHT1, AHT2} and  Floquet 
\cite{Geen_96, Geen_97} theory.

Here, we introduce a powerful approach for the design of heteronuclear decoupling sequences
based on multiple rotating frame technique.

In the absence of chemical shifts, the most straightforward way to decouple two heteronuclear spins $I$ and $S$ is to just irradiate
with full power on resonance to spin $I$. The presence of large chemical shifts makes this not the best strategy as using a
CW irradiation, one creates an effective field that is not perpendicular to the coupling interaction and therefore part of the
coupling interaction, parallel to this field is not averaged. We show that it possible to design a multiply modulated
rf-field, whose effect is best understood by performing a sequence of coordinate transformations, {\it specific to each chemical shift}.
Each transformation reduces the chemical shift interaction and averages part of the coupling until the chemical shifts are made
arbitrarily small. Each transformation demodulates a component of the multiply-modulated rf-field to a static component. The subsequent
transformation is in the rotating frame defined by new static component and the residual chemical shift. We show that by increasing the
number of modulations in the decoupling field, we can significantly improve the decoupling. Methods to make this method robust to
rf-inhomogeneity are discussed.

\section{Heteronuclear Decoupling}

Consider two heteronuclear spins $I$ and $S$. The Hamiltonian of the spin system takes the form

\begin{equation}
\label{eq:IS}
H_0 = \omega_I I_z + 2 \pi J I_z S_z + \omega_S S_z,
\end{equation}

where $\omega_I$ and $\omega_S$ are the chemical shifts of spin $I$ and $S$. Assuming
spin $S$ is being observed, we can without loss of generality take the chemical shift
of the spin $S$ to be zero and denoting $\omega_I = \omega_0$, consider the natural
Hamiltonian

\begin{equation}
\label{eq:control.0.r}
H_0 = \underbrace{\omega_0 I_z}_{H_{cs}} + \underbrace{2 \pi J I_z S_z}_{H_c},
\end{equation}

where $\omega_0 \in [-c_0, c_0]$. Now, consider the rf-Hamiltonian

\begin{equation}
\label{eq:control.1}
H_0' = w_0 I_x + (w_1 \sin \upsilon_1 t + w_2 \cos \upsilon_1 t  \sin \upsilon_2 t + \dots ) I_y.
\end{equation}

A more general form of the rf-Hamiltonian is then

\begin{equation}
\label{eq:control.2}
H_0' = w_0 I_x + ( \underbrace{\sum_{k=1}^{N} w_k \prod_{j=1}^{k-1} \cos \upsilon_j t \ \sin \upsilon_k t}_{A_0(t)}  )I_y .
\end{equation} where $\upsilon_k > \upsilon_{k+1}$. We adopt the notation $\bar{w}_k = 2^{-k} w_k$, and rewrite

\begin{equation}
\label{eq:frame.Hamiltonian2}
H_0 = \tilde{\omega} I_{z_1(\omega)} + A_0(t) I_y +  2 \pi J I_z S_z,
\end{equation} where $z_1(\omega)$ is the unit vector along the direction $\omega \hat{z} +  w_0 \hat{x}$ and define $\theta_1(\omega) = \tan^{-1}(\frac{w_0}{\omega})$, the spread of
frequencies $\tilde{\omega} \in [w_0, \sqrt{w_0^2 + c_0^2}]$. Note, $z_1(\omega)$, is different for each $\omega$. Now, by choosing $\upsilon_1$ in the first transformation as exactly
the center of this spread and transforming into a frame rotating around $I_{z_1(\omega)}$, with
frequency $\upsilon_1$, we get the Hamiltonian

\begin{equation}
\label{eq:frame.Hamiltonian3}
H_1 = \underbrace{(\tilde{\omega} - \upsilon_1)}_{f_1(\omega)} I_{z_1(\omega)} +  2 \pi J \cos \theta_1(\omega) I_{z_1(\omega)} S_z + \underbrace{\bar{w}_1 I_{x_1} + A_1(t) I_y}_{H_1'} + H_1''(t) + H_1''',
\end{equation}where $H_1'$ is the demodulated part of the Rf-Hamiltonian $H_0'$, in the interaction frame of
$I_{z_1(\omega)}$, that resembles $H_0'$ by design. $H_1''(t)$ and $H_1'''(t)$ are the fast oscillating parts of the rf and coupling Hamiltonian that we ideally want to average out and we neglect these terms for now. $H_1'''(t)$ is simply the part of the coupling perpendicular to the effective field direction  $I_{z_1(\omega)}$, that oscillates with frequency $\upsilon_1$.
The new frequency $f_1(\omega) \in [-c_1, c_1]$, where
$c_1 < c_0$. $A_1$, $H_1''$ and $H_1'''$ are written in their general form below.
The system obtained after first coordinate transformation has the desired
feature that the ratio of chemical shift spread to Rf-strength
$\alpha_1 = \frac{c_1}{\bar{w}_1}$ is reduced over $\alpha_0 = \frac{c_0}{w_0}$ for the original system. For example, if
$c_0 = w_0$, then $\frac{c_1}{\bar{w}_1} = \sqrt{2}-1 $.

We can now iterate the above construction. We go into successive rotating frames around axis $I_{z_k(\omega)}$ with frequency $\upsilon_k$.
Unit vectors $(x_k(\omega), y_k(\omega), z_k(\omega))$ define the $k^{th}$ frame, where we suppress the argument $\omega$ subsequently. $f_k(\omega)$ is the
chemical shift in the $k^{th}$ rotating frame, starting with $\omega$ in $H_0$. $c_k$ represents the limit of the chemical shifts in the $k^{th}$
coordinate frame and $\bar{w}_k = 2^{-k} w_k$ is the strength of the rf-field along the direction $x_k$.

\begin{equation}
\label{eq:Heq}
H_k =  f_k(\omega) I_{z_k(\omega)} + \frac{w_k}{2^k} I_{x_k} + 2 \pi J_k  I_{z_k}S_z + A_k(t) I_y + H_k''(t) + H_k'''(t),
\end{equation}

\begin{equation}
\label{eq:Aeq}
A_k(t) = \frac{1}{2^k} \{ \sum_{m=k+1}^{n} w_m \prod_{i=k+1}^{m-1}\cos(\upsilon_i t) \sin(\upsilon_m t) \}I_y,
\end{equation}

\begin{equation}
\label{eq:Heq''}
H_k''(t)=  \exp(i 2 \upsilon_k I_{z_k} t) (- \frac{w_k}{2^k} I_{x_k} +  A_k(t) I_y )\exp(-i 2 \upsilon_k I_{z_k} t), \end{equation}

\begin{equation}
\label{eq:Heq'''}
H_k'''(t)= J_k \exp(i \upsilon_k I_{z_k} t) I_{x_k}S_z \exp(-i \upsilon_k I_{z_k} t)
\end{equation}

\begin{equation}
\label{eq:Req}
c_{k+1} = \frac{\sqrt{c_k^2 + \bar{w}_k^2} - \bar{w}_k}{2} ; \ \ \  \upsilon_{k+1} = \frac{\sqrt{c_k^2 + \bar{w}_k^2} + \bar{w}_k}{2}; \ \ \ \tan \theta_k(\omega) = \frac{\bar{w}_k}{f_k(\omega)}
\end{equation}

\begin{equation}
\label{eq:JJ}
J_k = \prod_{k=1}^n \cos \theta_k(\omega) J.
\end{equation}

As discussed in the next section the iterated relations on $c_k$  insures that the ratio $ \alpha_k = \frac{c_k}{\bar{w}_k} $
is decreasing  and $\theta_k \rightarrow \frac{\pi}{2} $. This ensures that $J_k \rightarrow 0 $.

\section{Scaling}

To fix ideas, we choose $w_{k} = w_0$, i.e.,  $\bar{w}_k = 2^{-k} w_0$. From equation (\ref{eq:Req}), we have the relation,

\begin{equation}
\label{eq:alpha1.1}
\alpha_{k+1} = \sqrt{1 + \alpha_{k}^2} -1  < \alpha_{k}.
\end{equation}This ensures that $\alpha_k$ is decreasing. We explore two limits, for  $\alpha_k \ll 1 $,
\begin{equation}
\label{eq:alpha1.2}
\frac{\alpha_{k+1}}{\alpha_k} \sim   \frac{\alpha_k}{2}.
\end{equation} $\alpha_k \gg 1 $, we have

\begin{equation}
\label{eq:alpha.3}
\alpha_{k+1} \sim  \alpha_k - 1.
\end{equation}

Since $\alpha_k \rightarrow 0$, we have $\theta_k \rightarrow \frac{\pi}{2}$. This ensures that $J_k \rightarrow 0$.

We evaluate the root mean square amplitude for this rf-field as $N \rightarrow \infty$

\begin{equation}
\label{eq:rootmean}
A_{eff} = w_0 \sqrt{\sum_{k=1}^{N} 2^{-k}} \sim \sqrt{2} w_0.
\end{equation}The ratio $\frac{2 \upsilon_{k}}{\bar{w}_{k}}$ describes how well the oscillating component

\begin{equation}
\bar{w}_k \exp(j 2 \upsilon_k I_{z_k} ) I_{x_k} \exp(-j 2 \upsilon_k I_{z_k} ),
\end{equation}is averaged. This ratio

\begin{equation}
\frac{2 \upsilon_{k}}{\bar{w}_{k}} =  2 ( \sqrt{\alpha_{k-1}^2 + 1} + 1 ) > 4.
\end{equation}

\begin{equation}
\label{eq:upsilon}
\frac{\upsilon_{k}}{\upsilon_{k+1}} =  \frac{\bar{w}_k}{\bar{w}_{k+1}} \frac{( \sqrt{\alpha_{k}^2 + 1} + 1 )}{(\sqrt{\alpha_{k+1}^2 + 1} + 1 )}>2
\end{equation}

\section{Non-Resonant Conditions}


We now check whether all the oscillating terms captured by Hamiltonians $H_k''$ and $H_k'''$ that were neglected result in an effective coupling. If this were to happen; in the modulation frame of the rf-field and chemical shifts, we will see a net coupling evolution. Therefore, we evaluate the evolution of the couplings in the modulation frame defined by the Hamiltonian
$H_m = H_{cs} + H_{rf}$. The evolution of this frame takes the form

$$ U(\omega, t) = \exp(-i \upsilon_1 I_{z_1(\omega)} t) \dots \exp(-i \upsilon_k I_{z_k(\omega)} t) \dots \exp(-i \upsilon_n I_{z_n(\omega)} t) \Theta_n(\omega, t), $$

Where

\begin{equation}
\label{eq:coordinate.transform1}
\dot{\Theta}_n(\omega) = -i \{ f_n(\omega) I_{z(\omega)} + \tilde{H}(t) \} \Theta_n(\omega),
\end{equation}such that $| f_n(\omega) | \ll c_0 $. Here $f_n(\omega)$ represents the chemical shift in the $n^{th}$
frame. Where

\begin{equation}
\label{eq:residual.Hamiltonian}
\tilde{H}(t) = \sum_{k=1}^{N-1} V_{k+1}^{\dagger} H_k'' V_{k+1}
\end{equation}where

\begin{equation}
\label{eq:kunitary}
V_{k}(t) = \prod_{j=k}^N \exp(-i \upsilon_j I_{z_j} t).
\end{equation}

In this notation,

\begin{equation}
\label{eq:kunitary1}
 U(\omega, t) = V_1(t) \Theta_n(\omega, t).
\end{equation}

Where

\begin{equation}
\label{eq:Peano}
\Theta_n(\omega, t) = I + \int_0^t  \tilde{H} (\tau) d \tau + \int_0^t \int_0^{\tau} \tilde{H}(\tau) \tilde{H} (\sigma) d \tau d \sigma + \dots .
\end{equation}

Then in the frame of the Hamiltonian $H_m(t)$, the evolution takes the form

\begin{equation}
\label{eq:no-resonance.condition}
I - i 2 \pi J \int_0^t V_1'I_zS_z V_1 d \tau - 2 \pi J \int_0^t \underbrace{[V_1' I_zS_z V_1 , \int_0^{\tau}\tilde{H}(\sigma) d \sigma ]}_{J(\tau)} d \tau + \dots ...
\end{equation}where, the first integral is averaged out in the subsequent frames as
$\alpha_k \rightarrow 0$ and $\upsilon_k > 2 \upsilon_{k+1}$ in Eq. (\ref{eq:upsilon}) prevents generation of static components from the oscillating parts. We evaluate the second integral in the series above. If $J(\tau)$ has any static components this
would reflect residual couplings in the system. We use
the notation $H_k''(f)$ to denote the discrete set of frequencies present in oscillating Hamiltonian $H_k''(f)$.
The discrete frequencies is the set

\begin{equation}  \{ \exp(-i \omega I_\alpha) H_k'' \exp(i \omega I_\alpha) \}(f) = \{ \omega \pm H_k''(f) , \pm H_k''(f) \}
\end{equation}

$$ H_k''(f) = a_k \upsilon_k + \sum b_k \upsilon_{k+1}, a_k \in \{ \pm 2 \}, b_k \in\{ 0, \pm 1, \pm 2 \}  $$

Similarly, we have

$$ V_1' I_zS_z V_1 (f) = \sum_{k=1}^N c_k \upsilon_k, $$ where  $c_k \in \{ 0, \pm 1 \}$.
We compute the overlap of the two set of frequencies, we find the smallest value of $|\Delta|$ satisfying

\begin{equation}
\label{eq:gap}
\sum_{j=1}^{k-1} c_j  \upsilon_j + a_k \upsilon_k +  \sum_{j=k+1}^{N}  d_j  \upsilon_j = \Delta
\end{equation}

where  $d_j \in \{ 0, \pm 1 , \pm 2, \pm 3\}$. If $|\Delta| > 0$, we avoid a resonance condition.

\subsection{Simulation and Experiments}

\begin{figure}[ht]
\begin{center}
\includegraphics[scale=.4]{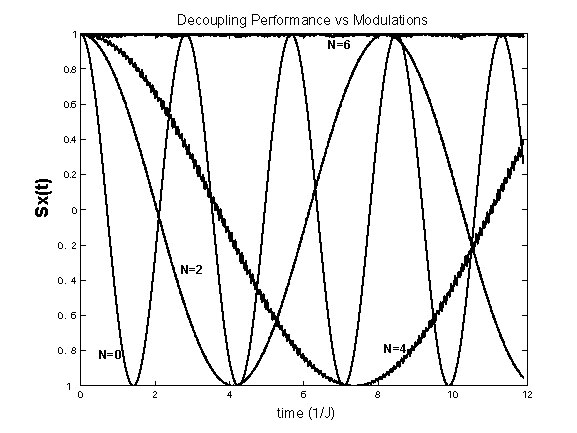}
\includegraphics[scale=.4]{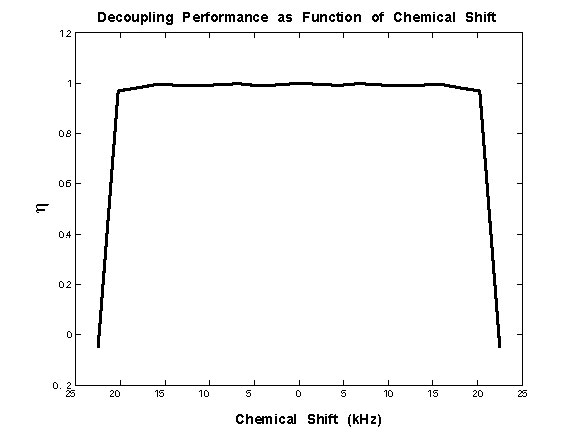}
\caption{The left figure shows how the coherence $S_x(t)$ evolves as a function of time in units of $J^-$ as $N$
the number of modulations in the Rf-field are increased. The simulations correspond to $J=140$ Hz, $A_{eff} \sim 6.25$ kHz and
$c_0 = 22.5$ kHz corresponding to $\frac{c_0}{A_{eff}} = 3.6$ and the chemical shift value $\frac{\omega_0}{A_{eff}} = 1$. The right figure shows the
decoupling efficiency $\eta = \frac{1}{T}\int_0^T S_x(t) dt$, as function of $\omega_0$. Here $T \sim \frac{12}{J} \sim 85$ ms.}
\label{fig:decoupling.perform}
\end{center}
\end{figure}

\begin{figure}[h]
\begin{center}
\includegraphics[scale=.4]{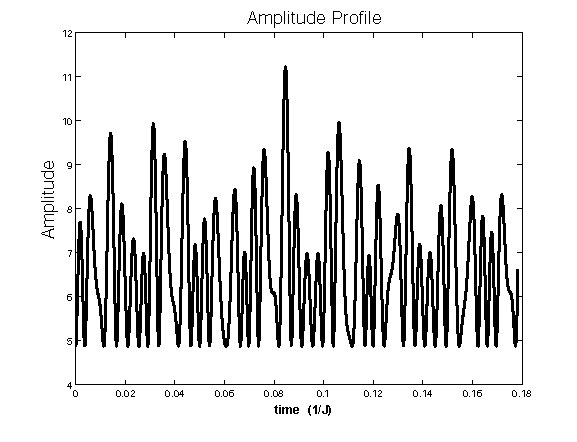}
\includegraphics[scale=.4]{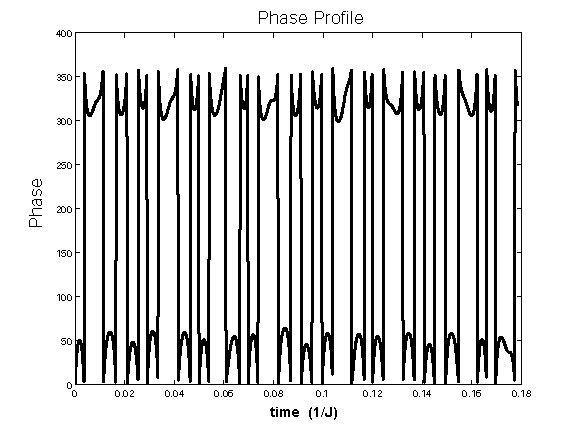}
\caption{The top figure shows the Amplitude (in kHz) of the rf-field as a function of time in units of $J^{-}$,
corresponding to $N=6$ modulations, with $w_0 = 4.8$ kHz. This corresponds to $A_{eff} = 6.25$ kHz and the maximum amplitude
$A_{max} = 12$ kHz. The bottom
panel shows the phase (in degrees) as function of time.}
\label{fig:amp-phase}
\end{center}
\end{figure}

Simulations were carried for a carbon-proton (IS) spin system, with a coupling constant $J = 140 Hz$ in Eq. (\ref{eq:IS}).
The Carbon chemical shift range was $c_0 (2\pi)^{-1} = 22.5$ kHz, which corresponds to a 200 ppm Carbon chemical shift at
900 mHz proton frequency. At $6.25$ kHz rf-power, the $\frac{\pi}{2}$ pulse corresponds to 40 $\mu s$.
We choose this as our $A_{eff}$ in the MODE sequence, so that $\frac{c_0}{A_{eff}} = 3.6$. We choose $w_k = w_0$,
and calculate $w_0$ that corresponds to $A_{eff} \sim 6.25$ kHz. For $N=6$, this corresponds to
$w_0 = 4.8$ kHz. Fig. \ref{fig:amp-phase} shows the amplitude and phase profile of the MODE sequence. Fig. \ref{fig:decoupling.perform}
shows evolution of initial magnetization $S_x(0)=1$ under the MODE sequence for various values of $N$ for a period of $\sim 12 J^{-}$.
Fig. \ref{fig:decoupling.perform} also shows the decoupling efficiency,

\begin{equation} \eta = \frac{1}{T}\int_0^T  S_x(t) dt,
\end{equation}for various offsets.

\begin{figure}[ht]
\begin{center}
\includegraphics[scale=.4]{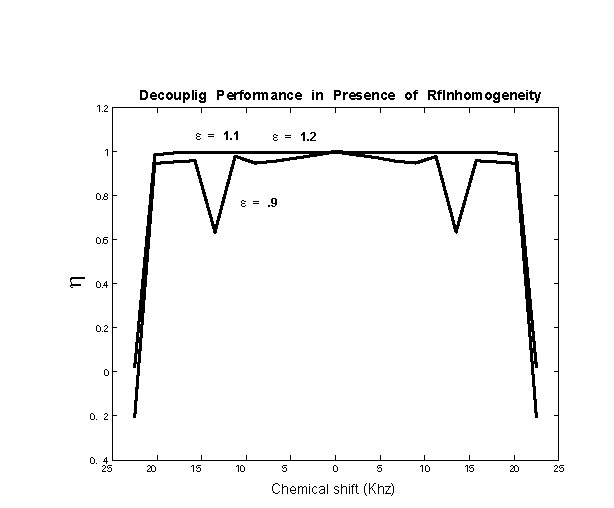}
\caption{ The above figure shows the
decoupling efficiency $\eta$, as function of $\omega_0$. Here $T \sim \frac{12}{J} \sim 85$ ms for rf-inhomogeneity values corresponding to $\epsilon = .9$,
$\epsilon = 1.1$ and $\epsilon = 1.2$ with $\delta = -.1$, $\delta=.1$ and $\delta = .2$ respectively. The mode sequence is designed with $N=6$, $\frac{c_0}{w_0} = 3.6$
and $c_0 = 22.5$ kHz.}
\label{fig:decoupling.inhomo}
\end{center}
\end{figure}

\begin{figure}[h]
\begin{center}
\includegraphics[scale=.4]{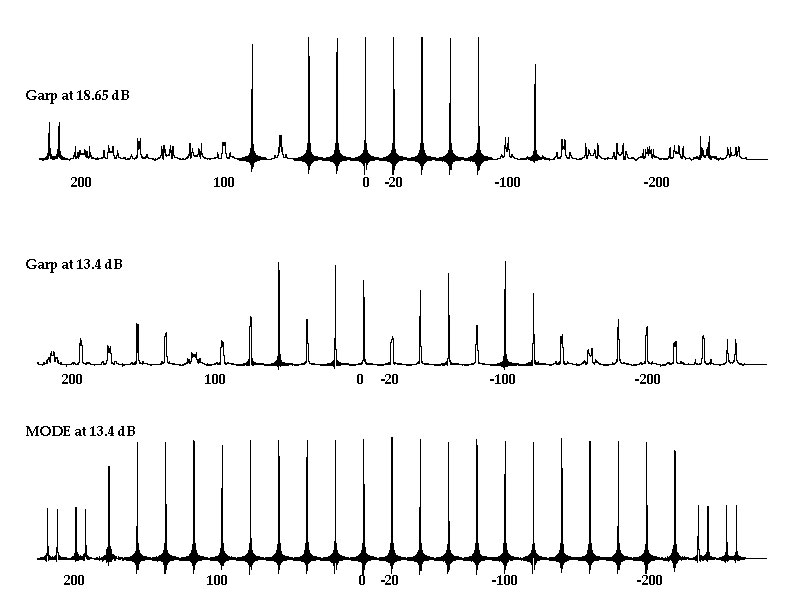}
\includegraphics[scale=.4]{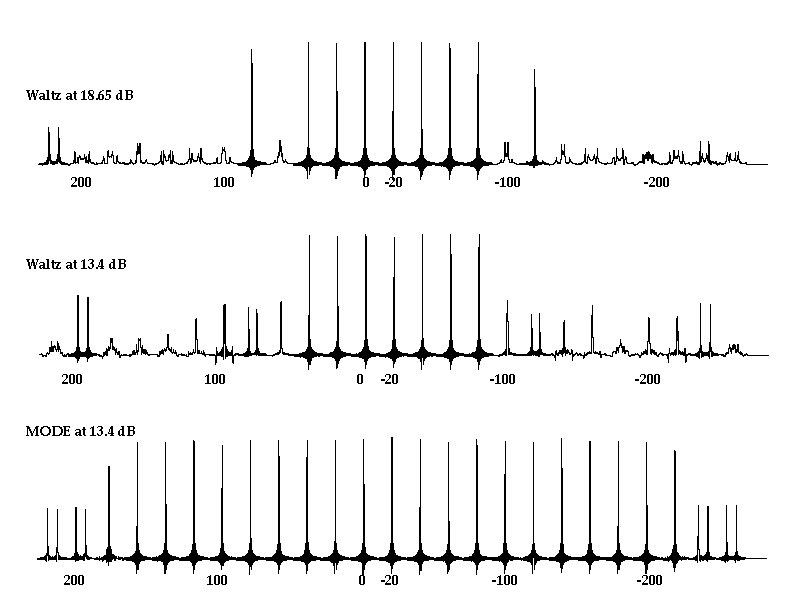}
\end{center}
\caption{The above figures show the experimental proton spectra obtained at 500 mHz for a
13C-iodomethane sample,
dissolved in D-chloroform (D, 99.8\%,Cambridge Isotope Laboratory, Inc.)
The rf-field corresponds exactly to the simulation parameters with $N=6$.
The power level of $13.4$ dB corresponds to a 40 $\mu$s $\frac{\pi}{2}$ pulse. i.e, a $6.25$ kHz Rf-field. The $A_{eff}$ for the MODE
decoupling sequence is at $6.25$ kHz. }
\label{fig:exp}
\end{figure}

\begin{figure}[h]
\begin{center}
\includegraphics[scale=.6]{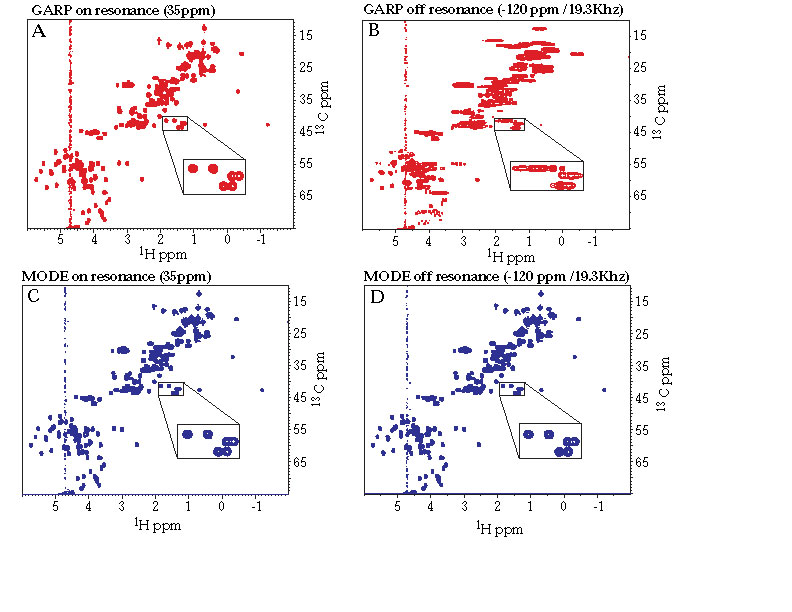}
\end{center}
\caption{The above figures show the HSQC spectra on a sample of the protein GB1 obtained at 500 mHz . In the indirect dimension, the proton is decoupled by a refocusing $\pi$ pulse. In the direct dimension, the carbon is decoupled by application of GARP and MODE decoupling sequences with mean rf-amplitude of $\sim 3.68$ kHz and $6.25$ khz respectively.
The MODE sequence used is the same described earlier in the section. A plasmid pET30-a containing the gene coding for the HIS-tagged GB1 protein (59 amino acid immunoglobulin binding domain from streptococcal protein G) was transformed into E. coli BL21(DE3) cells for protein expression. A 15N/13C uniformly labeled sample was prepared by overexpression in M9 minimal media containing 13C glucose and 15NH4Cl. The protein was initially purified using a Nickel resin (Qiagen). The HIS tag was later removed with TEV protease and the resulting GB1 was further purified on a FPLC using a Sephadex column. The sample was concentrated to a final concentration of 1 mM in phosphate buffer (20 mM, pH = 6.5) with 150 mM NaCl and 100 µM EDTA.
}
\label{fig:exp1}
\end{figure}

The parameters of the MODE sequence for $N=6$, $w_0 = 4.8$ kHz and $c_0 = 22.5$ kHz are
$c_1=9.1$, $c_2=3.5$, $c_3=1.23$, $c_4=.38$, $c_5=.09$ and $c_6=.01$ kHz. The ratio $\frac{c_0}{\bar{w}_0} = 4.6$,
$\frac{c_1}{\bar{w}_1} = 3.7$, $\frac{c_2}{\bar{w}_2} = 2.85$, $\frac{c_3}{\bar{w}_3} = 2.02$, $\frac{c_4}{\bar{w}_4} = 1.26$,
$\frac{c_5}{\bar{w}_5} = .61$ and $\frac{c_6}{\bar{w}_6} = .17$. In the last frame
the ratio of chemical shift to effective control is significantly reduced.  The smallest frequency $|\Delta|$ in Eq. (\ref{eq:gap}) is $197.3$ Hz
which avoids resonance.

Fig.(\ref{fig:exp}) shows the experimental spectra obtained on a system of methyl Iodide with MODE sequence. The experimental
parameters of the system are the same as in simulations. Comparison of conventional decoupling
sequences with MODE shows it is much broadband for same root mean square rf-power. Fig.(\ref{fig:exp1}) shows the 2D HSQC
spectra of the protein GB1, comparing GARP and MODE decoupling sequence on $^{13}C$ during direct detection.

The figure (\ref{fig:decoupling.inhomo}) shows the performance of the MODE decoupling sequence as function of rf-inhomogeneity,
which is captured by the parameter $\epsilon = (1 + \delta)$, such that the actual amplitude of the rf field is $\epsilon A $ where
$A$ is the nominal amplitude.

In the absence of inhomogeneity,
for all $\omega \in [-c_0, c_0]$, $f_k(\omega) \leq f_k(c_0)$ and the ratio $\frac{c_k}{\bar{w}_k}$ is constantly decreasing. In the presence of
rf-inhomogeneity, for our simulation with
$\delta = .1$ and $\delta = .2$, it is observed that $\frac{f_k(c_0)}{\bar{w}_k}$ begins to increase. For $\delta = -.1$, we observe that
$f_k(\omega) \not\le f_k(c_0)$. The performance of the MODE sequence in presence of rf-inhomogeneity for $\delta > 0$ is comparable to the ideal case.
For $\delta = -.1$ a small set of offsets are affected by rf-inhomogeneity. For these offsets, we find that the angle
$\prod_k \cos(\theta_k)$ in $\ref{eq:JJ}$ is the largest, indicating residual coupling in the system.

The following section discusses techniques to design in presence of rf-inhomogeneity.

\section{Discussion and Conclusion}

In the presence of rf-inhomogeneity, we can redefine our iterative procedure for computing the modulation frequencies
$\upsilon_k$. In presence of inhomogeneity, the spread of effective shifts after the first frame transformation
is from $[(1-\delta)w_0 , \sqrt{(1 + \delta)^2 w_0^2 + c_0^2} ]$, the left limit corresponds to zero offset and smallest Rf-amplitude.
The right limit corresponds to largest Rf-amplitude and chemical shift. We choose $\upsilon_1$ as center of this spread. Following this
we obtain that

\begin{equation}
\upsilon_{k+1} = \frac{\sqrt{c_k^2 + \bar{w}_k^2(1 + \delta)^2}  + \bar{w}_k (1 - \delta)}{2}  ;\ \ c_{k+1} = \frac{\sqrt{c_k^2 + \bar{w}_k^2(1 + \delta)^2}  - \bar{w}_k (1 - \delta)}{2}.
\end{equation}

This gives

\begin{equation}
\alpha_{k+1} = \sqrt{\alpha_k^2 + (1 + \delta)^2} - (1 - \delta).
\end{equation}

Note

\begin{equation}
\sqrt{\alpha_k^2 + (1 + \delta)^2} - (1 - \delta) \geq \alpha_k
\end{equation} with equality when $$ \alpha_k = \alpha_p = \frac{2 \delta}{1 - \delta}. $$

The $\alpha_k$ values decrease until they reach $\alpha_p$.  For $\delta = .1$, we have $\alpha_p \sim .22$, which is good enough as the ratio of chemical shift to Rf-strength is small. More generally, taking $w_{k+1} = \frac{w_k}{2 g_k}$, we can move this fixed point
close to zero.

In this paper, we introduced the multiple rotating field technique as a means to design broadband heteronuclear decoupling sequences in Solution NMR. It is important to point out that the MODE sequences for the case when $N=1$, simply reduce to the well known TPPM decoupling pulse sequence \cite{TPPM}. Also see \cite{Spinal-64, gan, eden, Yu, Takegoshi, XiX}. The main contribution of this paper lies in realizing the importance of the ratio $\alpha_k$ and showing that by adding  extra modulations in the rf-field and successively transforming into rotating frames, we can significantly reduce the ratio $\alpha_k$ and
improve the decoupling performance. The technique is also expected to find use in designing decoupling pulse sequences in Solid State NMR and design of various excitation, inversion and mixing sequences, where spread of chemical shifts can simply be removed by transforming into a suitable $K$ frame.

\section{Acknowledgement}

Authors would like to thank Prof. Steffen J. Glaser for helpful disciussions on the subject
and pointing out a complete set of references \cite{tracking}. N. Khaneja will like to
acknowledge NSF-0724057, ONR 38A-1077404 and AFOSR FA9550-05-1-0443 for supporting this work.

\end{document}